\documentclass[12pt]{article}
\usepackage{amssymb}
\usepackage{amsmath}
\usepackage{graphicx}

\numberwithin{equation}{section}

\begin{document}
\baselineskip=17.5pt

\begin{titlepage}

%\begin{flushright}
%{\small xxx}\\[-1mm]%
%{\small xxx}%
%\end{flushright}

\begin{center}
\vspace*{10mm}

{\large\bf Power counting renormalizability of scalar theory
\\
in Lifshitz spacetime}%
\vspace*{12mm}

Takayuki Hirayama\footnote{taka.hirayama@gmail.com}
\vspace{4mm}

{\it Fukashi High School, Arigasaki 3-8-1, Matsumoto, Nagano 390-8603, 
Japan}\\%
\vspace*{12mm}

\begin{abstract}\noindent%
We analyse the power counting renormalizability of scalar theory in Lifshitz spacetime in $D+2$ dimensions. We show the spectral dimension becomes $2+(D/z)$ ($z$ is the critical exponent) after integrating out the radion field. We comment on the AdS/CFT correspondence, and on how to avoid the Lifshitz singularity by flowing into AdS spacetime in the infrared. We also comment on the quantum gravity in Lifshitz spacetime.
\end{abstract}

\end{center}

\end{titlepage}

\newpage

%%%%%%%%%%%%%%%%%%%%%%%%%%%%%%%%
%%%%%%%%%%%%%%%%%%%%%%%%%%%%%%%%
%%%%%%%%%%%%%%%%%%%%%%%%%%%%%%%%
\section{Introduction}

General relativity is not renormalizable, if quantized, and higher loops induce more and more severe UV divergences~\cite{'tHooft:1974bx}. Since the gravitational coupling has the mass dimension $2-n$ in $n$ dimensional spacetime, $L$ loop graphs induce divergences with the power of $\Lambda_{UV}^{(n-2)L+2}$ where $\Lambda_{UV}$ is the UV cutoff in more than two dimensions. For example in four dimensional spacetime ($n=4$), the one loop matter graphs induce upto dimension four operators with an appropriate power in the cutoff, i.e. the cosmological constant, Einstein Hilbert term and quadratic terms in Riemann tensor ($R^2$, $R_{ab}R^{ab}$ and $R_{abcd}R^{abcd}$). As long as some symmetries or identities, such as supersymmetry and Bianchi identities, do not forbid such terms, they are induced. Since General relativity does not have quadratic terms in the action, we can not cancel the divergences in front of the quadratic terms in Riemann tensor and the quantum theory is not renormalizable. 

If we start from the higher derivative gravity in four dimensions\footnote{In four dimensions, $\sqrt{-g} \{ R^2 -4 R_{ab}R^{ab} +R_{abcd}R^{abcd} \}$ becomes total derivatives and then $R_{abcd}R^{abcd}$ can be rewritten by $R^2$ and $R_{ab}R^{ab}$.},
\begin{align}
 S &= \!\! \int \! \! d^4 \! x \sqrt{-g} \left[ M^{n-2} ( R-2\Lambda) +\alpha R^2 +\beta R_{ab}R^{ab} + \gamma R_{abcd}R^{abcd} \right],
\end{align}
the gravitational coupling has the mass dimension zero and we can cancel the divergences using the coefficients in front of quadratic terms in Riemann tensor ($\alpha$, $\beta$ and $\gamma$). Thus the theory is renormalizable~\cite{Stelle:1976gc}, however, the graviton propagator behaves $1/k^2(k^2-m^2)$ where $m$ is computed from the action and this has additional poles, in addition to the one for massless spin two graviton. And this corresponds to a massive spin two graviton and massive spin zero graviton. The massive spin two graviton has a negative kinetic energy, thus a ghost particle, and then the unitarity and/or stability of the flat spacetime is lost. Therefore this non renormalizability remains as one of important problems in quantum gravity and this leads many thoughts in the concept of spacetime at the short distance.

Horava~\cite{Horava:2009uw} recently proposed an idea that Lorentz symmetry is broken at UV, but a theory possesses an invariance under the anisotropic rescaling (parametrized by $\lambda$) with a dynamical exponent~$z$,
\begin{align}
 t \rightarrow \lambda t , \hspace{3ex}
 x^i \rightarrow \lambda^{1/z} x^i.
\end{align}
A theory with this invariance is a Lifshitz scalar theory~\cite{lifshitz} whose action is
\begin{align}
 S &= \!\! \int d t d^D\! x \left[ (\partial_t \phi)^2 - ( \partial_i^{\: z} \phi)^2 \right].
\end{align}
Then the propagator is given $1/(\omega^2 -\vec{k}^{2z})$ where $\omega$ and $\vec{k}$ are the momentum along time and spacial directions respectively. Then the propagator does not contain any additional poles and dumps quickly along large spacial momentum ($z>1$) resulting that the spectral dimension becomes $1+(D/z)$ at UV instead of $1+D$~\cite{Horava:2009if}. Therefore the UV divergences are suppressed, and Horava constructed an action of renormalizable quantum gravity with this invariance. 

On the other hand, there is a spacetime, Lifshitz spacetime~\cite{Koroteev, Kachru:2008yh, Taylor:2008tg}, whose isometries match with the anisotropic rescaling,
\begin{align}
 ds^2 &= L^2 \left( -\frac{dt^2}{r^2} + \frac{dx_D^2}{r^{2/z}} + \frac{dr^2}{z^2r^2} \right).
\end{align}
This metric is invariant under the change $t\rightarrow \lambda t$, $x^i\rightarrow \lambda^{1/z}x^i$ and $r\rightarrow \lambda r$. This metric can be a solution of, for example, Einstein gravity with massive gauge fields or higher derivative gravity theories~\cite{Kachru:2008yh}. Then we expect that a theory ($L=\sqrt{-g} \: (\partial\phi)^2$) in Lifshitz spacetime shares the same improvement of UV divergences. However questions quickly arise. How do the higher spacial derivatives appear, since the action does not have higher spacial derivatives? Since the spacetime curvature is naively negligible at a short distance, how can the UV divergences be suppressed? These are the issues we discuss in this paper.

We study a scalar theory in Lifshitz spacetime, and analyse the propagator. We use Kaluza-Klein picture after compactifying $r$ direction by introducing a cutoff at large $r$, ($r=R$). This is because it is easier to study the UV behaviour of the propagator and we avoid the null singularity appears at $r=\infty$ in the metric~\cite{Kachru:2008yh}. We show the propagator becomes $1/(\omega^2-\vec{k}^{2z} -k_r^2)$, $k_r$ represents the momentum along $r$ direction, after integrating out the radion fields and the spectral dimension becomes $2+(D/z)$ at UV. Thus the UV behaviour is improved. We then discuss how the theory is modified when the metric flows into AdS metric toward $r=\infty$~\cite{Kachru:2008yh, Park:2012mn}. We show that the lower spacial derivatives $(\partial_i^{\: n}\phi)^2$, ($n=1,\cdots,z)$, are induced and the Lorentz symmetry is recovered in the infrared. This is consistent with the AdS/CFT correspondence~\cite{Maldacena:1997re}. We then give discussion on the renormalizability of quantum gravity in Lifshitz spacetime.

%%%%%%%%%%%%%%%%%%%%%%%%%%%%%%%%
%%%%%%%%%%%%%%%%%%%%%%%%%%%%%%%%
%%%%%%%%%%%%%%%%%%%%%%%%%%%%%%%%

\section{Scalar theory in the Lifshitz geometry}

We study a scalar theory in the Lifshitz geometry and the propagator in order to analyse the degree of UV divergence of general Feynman graphs. The action we use is a $D+2$ dimensional action,
\begin{align}
 S &= \!\! \int \!\! dt d^D \! x dr \sqrt{-g} \left[ -(\partial \phi)^2 -m^2 \phi^2 - y \phi^n \right].
\end{align}
The background geometry is the Lifshitz geometry~\cite{Koroteev, Kachru:2008yh, Taylor:2008tg},
\begin{align}
 ds^2 &= L^2 \left( -\frac{dt^2}{r^2} + \frac{dx_D^2}{r^{2/z}} + \frac{dr^2}{z^2r^2} \right).
 \label{102}
\end{align}
If we define a new radial coordinate $u=1/r^{1/z}$, the metric becomes a well known one,
\begin{align}
 ds^2 &= L^2 \left( -u^{2z}dt^2 + u^2dx_D^2 + \frac{du^2}{u^2} \right).
\end{align}
This metric can be a solution in Einstein gravity with massive gauge fields or higher derivative gravity theories. The Lifshitz metric has isometries under the anisotropic rescaling,
\begin{align}
 t \rightarrow \lambda t, \hspace{3ex} x^i \rightarrow \lambda^{1/z} x^i, \hspace{3ex}
 r\rightarrow \lambda r.
\end{align}
Substituting the metric into the action, the kinetic terms of the action $S_2$ becomes
\begin{align}
 S_2 &= \!\! \int \!\! dt d^D \! x dr \frac{L^D}{z r^{2+(D/z)}} \left[
 r^2 (\partial_t \phi)^2 - r^{(2/z)} ( \partial_i \phi)^2 -z^2r^2(\partial_r\phi)^2 -m^2L^2\phi^2
 \right]
 \\
 &= \!\! \int \!\! dt d^D \! x dr \frac{L^D}{z r^{(D/z)}} \phi \left[
 -\partial_t^2 + r^{-2+(2/z)} \partial_i^2 +z^2\partial_r^2 -Dzr^{-1} \partial_r -m^2L^2 r^{-2}
 \right]\phi.
\end{align}
Then the free part of equation of motion becomes
\begin{align}
 \omega^2\phi_{(\omega,\vec{k})}(r) &= \left[
 r^{-2+(2/z)} k^2 -z^2\partial_r^2 +Dzr^{-1} \partial_r +m^2L^2 r^{-2}
 \right] \phi_{(\omega,\vec{k})}(r),
 \label{128}
\end{align}
where we use the Fourier expansion (taking the real part),
\begin{align}
 \phi(t,x^i,r) &= \!\! \int \!\! d\omega d^D\! k \: \phi_{(\omega,\vec{k})}(r) e^{i(\omega t + k_ix^i)} .
\end{align}
For $k^2=\sum_i k_i^2 \neq 0$, we define $\widetilde{r}=k^zr$ and have
\begin{align}
 \omega^2 k^{-2z} \phi_{(\omega,\vec{k})}(\widetilde{r}) &=
 \left[
  \widetilde{r}^{-2+(2/z)}  - z^2 \partial_{\widetilde{r}}^2 +Dz\widetilde{r}^{-1} +m^2L^2\widetilde{r}^{-2}
 \right]\phi_{(\omega,\vec{k})}(\widetilde{r}).
\end{align}
It is seen that the spacial momentum $k_i$ appears only in the combination $\omega^2/k^{2z}$. Further we define $\phi_{(\omega,\vec{k})}(\widetilde{r}) = \widetilde{r}^{D/2z}\widetilde{\phi}_{(\omega,\vec{k})}(\widetilde{r})$, and we obtain
\begin{align}
 \left[ -\partial_{\widetilde{r}}^2 +V(\widetilde{r}) \right] \widetilde{\phi}_{(\omega,\vec{k})}(\widetilde{r})
 &= E \widetilde{\phi}_{(\omega,\vec{k})}(\widetilde{r}),
 \label{143}
 \\
 V(\widetilde{r}) &= \frac{D^2+2Dz+4L^2m^2}{4z^2} \frac{1}{\widetilde{r}^2} +\frac{1}{z^2} \frac{1}{\widetilde{r}^{2-(2/z)}},
 \\
 E&= \omega^2 k^{-2z} z^{-2}.
\end{align}
When $z=1$, this equation of motion reduces to the one in AdS$_{D+2}$ spacetime. Since the potential $V$ goes to zero at large $\widetilde{r}$, the energy $E$ is continuous and starts from zero.

It is good to use the Kaluza-Klein picture by compactifying $r$ direction with a cutoff at large $r$, $r=R$, in order to study the UV behaviour of the propagator as we will see shortly. Actually we have another reason to compactify $r$ direction which is that we avoid the null singularity at $r=\infty$ where tidal forces diverge~\cite{Kachru:2008yh}. We will come to this point later. 

The asymptotic solutions near $r=0$ and $r=\infty$ are
\begin{align}
 \chi_{U}^{\pm}(\widetilde{r}) &= \widetilde{r}^{ \frac{1}{2}\left( 1 \pm \sqrt{ 1 + (D^2+2Dz+4L^2m^2)/z^2 )} \right)  },
 \\
 \chi_{I}^{\pm}(\widetilde{r}) &= e^{\pm i \sqrt{E} \widetilde{r}} ,
\end{align}
and thus the mass $m$ should satisfy $1 + (D^2+2Dz+4L^2m^2)/z^2 \geq 0$ for having a normalizable mode. The normalizable modes should approach to $\chi_{U}^+(\widetilde{r})$ at $r=0$ and satisfy the even or odd boundary condition at the cutoff $r=R$, i.e. $\partial_r \phi|_{r=R}=0$ or $\phi(r)|_{r=R}=0$. Since we are interested in the UV behaviour of the propagator, it is not important whether we give the even or odd boundary condition. Since the action becomes
\begin{align}
 S_2 &= \!\! \int \!\! dt d^D \! x dr \: L^D z (E \widetilde{\phi} )^2 +\cdots,
\end{align}
we should take $\chi_U^+$ for a normalizable mode.

\begin{figure}[t]\begin{center}
  \includegraphics[width=14.0cm]{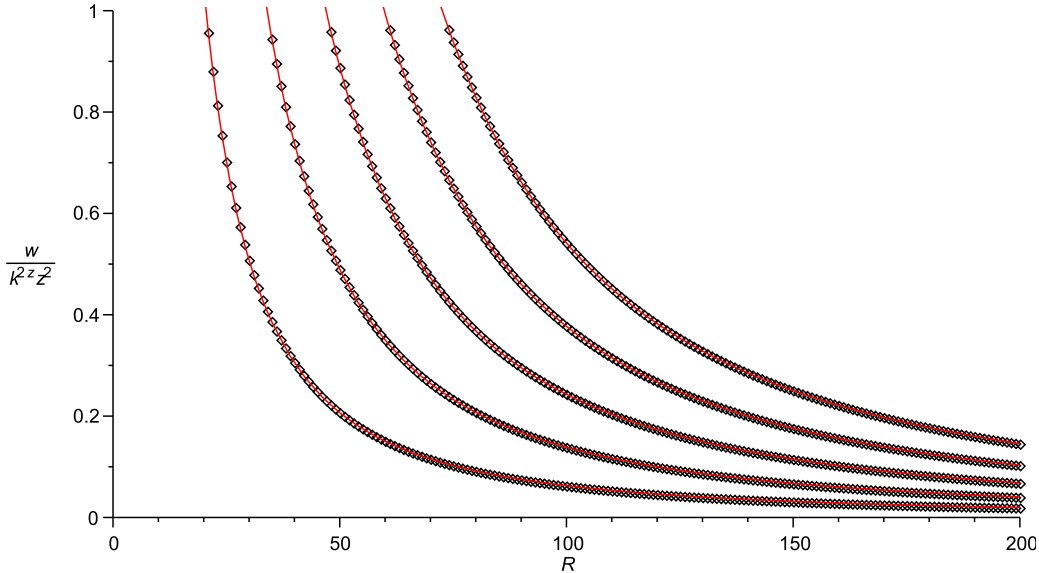}
\caption{$w^2/k^{2z}z^2$ for $R$.}
 \label{f1}
\end{center}\end{figure}

Now we compute the Kaluza-Kline modes along $r$ direction. Since the equation of motion is complicated, we rely on numerical analysis. We choose $D=3$, $z=4$, $m=0$, and the odd boundary condition at $r=R$ as a benchmark, and numerically solve the differential equation. We computed upto the fifth eigenmodes (see Fig.\ref{f1}).
%\begin{center} 
%\end{center}
Of course, the Kaluza-Klein modes continue infinitely. The values are found to be well fit with the following form, ($\widetilde{R}=k^zR$)
\begin{align}
 E_n  &= \frac{c_n}{\widetilde{R}^{2-(2/z)}} +\frac{d_n}{\widetilde{R}^2}, \hspace{3ex}
 \rightarrow
 \hspace{3ex}
 \omega_n^2 = \frac{c_n z^2}{R^{2-(2/z)}} k^2 +\frac{d_n z^2}{R^2},
 \label{178}
\end{align}
with
\begin{align}
 (c_1 z^2, ~d_1 z^2 ) &= (37.49,~ 251.18),&
 (c_2 z^2, ~d_2 z^2 ) &= (55.46,~ 830.42),
 \\
 (c_3 z^2, ~d_3 z^2 ) &= (70.76,~ 1721.85),&
 (c_4 z^2, ~d_4 z^2 ) &= (84.51,~ 2925.35),
 \\
 (c_5 z^2, ~d_5 z^2 ) &= (97.01,~4442.96).
\end{align}
From these numerical values, $c_n$ and $d_n$ are roughly fit with $c_nz^2\approx 4(n^{2-(2/z)}+6n^{1-(2/z)})$ and $d_n z^2 \approx 200 ( n^2 -(n/3) )$. Therefore this form is not an expected one from the anisotropic rescaling, $\omega^2=k^{2z}+k_r^2$ where $k_r$ represents the momentum along $r$ direction.

The $R$ dependence in~\eqref{178} is expected from the equation of motion in~\eqref{143} since the equation only depends on $\widetilde{r}$ and $\widetilde{r}^{2-(2/z)}$. Then after the Kaluza-Klein reduction and properly normalizing the fields, we obtain the following effective $D+1$ dimensional action for the kinetic terms,
\begin{align}
 S_{D+1} &= \!\! \int dt d^D \! x \sum_n \left[ (\partial_t \phi^{(n)} )^2
 - \frac{C_n}{R^{2-(2/z)}} (\partial_i\phi^{(n)} )^2 -\frac{D_n}{R^2} (\phi^{(n)} )^2
 \right],
\end{align}
where $C_n$ and $D_n$ are numerical factors whose $n$ dependence are $n^{2-2(/z)}$ and $n^2$ at large $n$. Then the propagator of each mode does not have the expected one, $1/(\omega^2-k^{2z}-m^2)$, but has an usual one, $1/(\omega^2 -k^2 -m^2)$ and thus we may conclude the UV behaviour is not improved although the theory has the invariance under the anisotropic rescaling. Here we should take into account the effects of the radion fields. The radion field is nothing but the component in the metric which is responsible for the anisotropic rescaling and thus corresponds to the shift of $R$. Then the equation of motion for the radion fields ends up the equation of motion for $R$ by treating $R$ as a field. The equation is then schematically written as
\begin{align}
 R^{2/z} \sim \left( \sum_m D_m (\phi^{(m)})^2 \right) \left/ \left( \sum_n C_n (\partial_i \phi^{(n)})^2 \right)  \sim
 N^{2/z} / (\partial_i)^2 \right.
\end{align}
where we introduced a cutoff $N$ for the number of eigenmodes and replaced $(\phi^{(n)})^2$ by the VEV. Then plugging this into the action, we have
\begin{align}
 S_{D+1} \sim \int \!\! d^{D+1}\! x \sum_n \left[
  (\partial_t \phi^{(n)})^2 -\widetilde{C}_n (\partial_i^{\: z} \phi^{(n)} )^2 
 \right],
\end{align}
where $\widetilde{C}_n \sim n^2/N^2$, (or $\sim n^{2-(2/z)}/N^{2-(2/z)})$ and thus $\widetilde{C}_n$ is small for small $n$ and becomes order one at large $n$. Although this computation is rough, this computation shows that the propagator dressed with the radion fields becomes $1/(\omega^2-k^{2z})$ as we expected. One may think that it is not justified to integrate out the radion field since it must be massless. The point is that the two point function of scalar field after summing up the Feynman graphs whose intermediate state contains the radion fields becomes the expected one (we do not have to really integrate out the radion fields).

In total, the propagator of the scalar field in $D+2$ dimensions at high energy behaves like
\begin{align}
 \frac{1}{\omega^2 -k^{2z} -k_r^2},
\end{align}
after taking into account the contribution of the radion fields, and $k_r$ represents the momentum along $r$ direction ($n^2/R^2\rightarrow k_r^2$). With this propagator, it is easy to compute the degree of UV divergence for a general Feynman graph. If we have a loop integration, we define $s_i=k_i^{\: z}$ and have
\begin{align}
 \int \!\! d\omega d^D\! k dk_r \frac{1}{\omega^2 -k^{2z} -k_r^2} &=
  \int \!\! d\omega d^D\! s dk_r z^{-D} \left(\prod s_i^{-1+(1/z)} \right)
  \frac{1}{\omega^2 -s^{2} -k_r^2}.
\end{align}
Then the superficial degree of UV divergence $\Gamma$ for $\phi^n$ scalar theory is
\begin{align}
 \Gamma &= \left\{ D+2-D(1-z^{-1}) \right\} L -2I = \left\{ 2+(D/z) \right\} L -2I,
\end{align}
where $L$ and $I$ are the number of loops and internal lines respectively. We then can see that the spectral dimension becomes $2+(D/z)$ at high energy. Using the relation $I=V+L-1$ and $nV=E+2I$ where $V$ and $E$ are the number of vertices and external lines respectively, we have
\begin{align}
 \Gamma &= \left\{ \frac{D}{z}\left(\frac{n}{2}-1\right)-2\right\} V -\frac{D}{2z}E +2+\frac{D}{z}.
\end{align}
Then $\phi^n$ theory is power counting renormalizable when $n\leq 2+(4z/D)$.

%%%%%%%%%%%%%%%%%%%%%%%%%%%%%%%%
%%%%%%%%%%%%%%%%%%%%%%%%%%%%%%%%
%%%%%%%%%%%%%%%%%%%%%%%%%%%%%%%%

\section{Discussions}

As discussed in the previous section, the propagator in ($D+2$) dimensional Lifshitz geometry with the exponent $z$ behaves like $1/(\omega^2-k^{2z}-k_r^2)$ at UV and the spectral dimension becomes $2+(D/z)$. One key point is the existence of radion field whose contribution to the two point function of scalar field should be taken into account. To show that, we introduced the cutoff at large $r$, $r=R$. One reason is that it is easier to study the UV behaviour of the propagator from the Kaluza-Klein picture. Another reason is that we avoid the null singularity appears at $r=\infty$ where it is computed that the tidal forces diverge. One idea of resolving the singularity is the Lifshitz geometry flows into the AdS geometry toward $r=\infty$. Such solutions have been constructed~\cite{Kachru:2008yh, Park:2012mn}, and $g^{ii}$ in~\eqref{102} is replaced by something like $g^{ii} \sim r^{2/z} +a_{z-1}r^{2/(z-1)} +\cdots +a_1 r^2$. At large (small) $r$, $r^2$ ($r^{2/z}$) becomes dominant in $g^{ii}$, and the geometry is the Lifshitz geometry at small $r$ and approaches to AdS geometry toward $r=\infty$. Then if we solve the equation of motion for eigenmode similar to the equation~\eqref{128}, we obtain $\omega^2 = ( c_z/R^{2-(2/z)} +c_{z-1}/R^{2-(2/(z-1))}+\cdots +c_1 ) k^2 + d/R^2$. Then again solving the equation of motion for the radion $R$ and plugging the solution in the action, we obtain lower spacial derivative terms $(\partial_i^{\: n} \phi)^2$, $n=1,\cdots, z$, in the action. Thus at low energy the theory recovers the $(D+1)$ dimensional Lorentz symmetry. This is consistent with AdS/CFT correspondence in which the radial coordinate $r$ corresponds to the energy scale in CFT side (A large $r$ corresponds to IR in CFT.). We notice that as long as the geometry is not exactly AdS, but is the Lifshitz geometry at $r=0$ and is approaching to AdS toward $r=\infty$, we have $1/k^{2z}$ behaviour anywhere in the bulk and the UV behaviour does not change. 

From the computation in this paper, we expect that the UV divergences in quantum gravity in the Lifshitz geometry are suppressed compared with those in the flat spacetime. It is true since the spectral dimension becomes $2+(D/z)$ at UV region~\cite{hirayama}. However since the spectral dimension is still larger than two, the quantum theory of gravity in the Lifshitz geometry is power counting non-renormalizable. In order to improve this situation, one may use supersymmetry, the Gauss-Bonnet terms or higher derivative terms. Since the gravitational coupling has mass dimension $-D/z<0$ (the mass dimension of time derivative is one), supersymmetry (which is softly broken) does not help. Some terms will be forbidden by supersymmetry, but some higher derivative terms are allowed by the symmetry and those are induced by loop graphs in the end with divergent coefficients. The Gauss-Bonnet term does not induce higher derivatives in the linearlized Einstein equation around the Lifshitz metric and does not change the behaviour of propagator along large $k_r$. One may expect that a special combination of higher derivative term induces only higher spacial derivatives, but does not induce higher time derivatives so that ghost particles do not appear. This is one of interesting directions for constructing a renormalizable quantum gravity.

We normally expect that the UV behaviour of propagator does not change in a curved spacetime from that in the flat spacetime, since at a short distance, the curvature of spacetime is negligible. However as we discuss in this paper, the UV behaviour does change. Therefore it is worth investigating more detail for finding new regularizing methods and a renormalizable quantum gravity theory.

%%%%%%%%%%%%%%%%%%%%%%%%%%%%%%%%
%%%%%%%%%%%%%%%%%%%%%%%%%%%%%%%%
%%%%%%%%%%%%%%%%%%%%%%%%%%%%%%%%

\subsection*{Acknowledgments}
\noindent
I thank my wife for the encouragement.

%%%%%%%%%%%%%%%%%%%%%%%%%%%%%%%%
%%%%%%%%%%%%%%%%%%%%%%%%%%%%%%%%
%%%%%%%%%%%%%%%%%%%%%%%%%%%%%%%%

\end{document}